\documentstyle[prl,aps,multicol,epsfig,epsf,psfig,array]{revtex}
\newcommand{\beq}{\begin{equation}}
\newcommand{\eeq}{\end{equation}}
\newcommand{\bqa}{\begin{eqnarray}}
\newcommand{\eqa}{\end{eqnarray}}

\begin{document}
\setlength{\baselineskip}{0.333333in} 
\tightenlines

\draft
\author{U. Al Khawaja$\;^1$, J.O. Andersen$\;^1$,
N.P. Proukakis$\;^{1,2}$,
and H.T.C. Stoof$\;^1$
}
\address{\it $^1$Institute for Theoretical Physics,
\\Utrecht University,
Leuvenlaan 4, 3584 CE Utrecht, The Netherlands
\\$^{2}$Foundation for Research and Technology Hellas, \\
Institute of Electronic Structure and Laser, P.O. Box 1527, Heraklion 71 110, Crete, Greece
\\ (\today)}

\title{Erratum: Low-dimensional Bose gases [Phys. Rev. A 66, 013615 (2002)]}

\maketitle

\begin{abstract}
\end{abstract}


In Ref~\cite{paper}, we presented a modified Popov theory that is capable
of calculating the quasicondensate density $n_0$ in arbitrary dimensions
and at all temperatures. When a condensate exists, i.e., in
three dimensions
at sufficiently low temperatures
and in two dimensions
at zero temperature, the condensate
density $n_c$ is defined by the off-diagonal long-range order
of the one-particle density matrix. In particular,
in the modified Popov theory
we have that
\bqa
n_{\rm c}&=&\lim_{{\bf x}\rightarrow\infty}n_0e^{-{{1\over2}
\left[\langle\hat{\chi}({\bf x})-\hat{\chi}({\bf 0})\right]^2\rangle}}\;,
\label{1}
\eqa
where $n_0$ satisfies the equations
\bqa
n&=&n_0+{1\over V}\sum_{\bf k}
\Bigg[
{\epsilon_{\bf k}-\hbar\omega_{\bf k}\over2\hbar\omega_{\bf k}}
+{n_0T^{\rm 2B}(-2\mu)\over2\epsilon_{\bf k}+2\mu}
+{\epsilon_{\bf k}\over\hbar\omega_{\bf k}} N(\hbar\omega_{\bf k}) \Bigg]\;,
\label{h1}\\
\mu&=&(2n-n_0)T^{\rm 2B}(-2\mu)\;,
\label{h2}
\eqa
and the phase fluctuations are determined by
\bqa
\langle\hat{\chi}({\bf x})\hat{\chi}({\bf 0})\rangle
&=&{T^{\rm 2B}(-2\mu)\over V}\sum_{\bf k}\Bigg[
{1\over2\hbar\omega_{\bf k}}
\left[1+2N(\hbar\omega_{\bf k})\right]
-{1\over2\epsilon_{\bf k}+2\mu}\Bigg]e^{i{\bf k}\cdot{\bf x}}\;.
\label{h33}
\eqa
Here $n$ is the total density of the gas, $T^{\rm 2B}(-2\mu)$ is the
two-body matrix at energy $-2\mu$, and $\mu$ is the chemical potential.
$\epsilon_{\bf k}=\hbar^2k^2/2m$ and
$\hbar\omega_{\bf k}
=\left[\epsilon_{\bf k}^2+2n_0T^{\rm 2B}(-2\mu)\epsilon_{\bf k}\right]^{1/2}$.
In Ref.~\cite{paper}, we assumed that the exponent in Eq.~(\ref{1})
vanishes in the limit ${\bf x}\rightarrow\infty$
and thus found that at zero temperature the
depletion of the condensate in two dimensions equals
\bqa
{n-n_0\over n}&=&{1\over4\pi}\left(1-\ln2\right)T^{\rm 2B}(-2\mu)\;,
\label{dep1}
\eqa
whereas in three dimensions it equals
\bqa
{n-n_0\over n}&=&
\left({32\over3}-2\sqrt{2}\pi\right)\sqrt{na^3\over\pi}\;,
\label{dep2}
\eqa
where $a$ is the scattering length and we have used
$T^{\rm 2B}(-2\mu)=4\pi a\hbar^2/m$.
A careful analysis shows, however,
that for a consistent calculation of the quantum depletion
the above assumption
is incorrect and we have to take the contribution from the
phase fluctuations into account.
In the limit ${\bf x}\rightarrow\infty$, we obtain
in two and three dimensions
\bqa
\left[\langle\hat{\chi}({\bf x})-\hat{\chi}({\bf 0})\right]^2\rangle
&=&-{1\over2\pi}\left(\ln2\right)T^{\rm 2B}(-2\mu)\;,
\label{corr1}
\eqa
and
\bqa
\left[\langle\hat{\chi}({\bf x})-\hat{\chi}({\bf 0})\right]^2\rangle
&=&
\left(2-\sqrt{2}\pi\right)\sqrt{na^3\over\pi}\;,
\label{corr2}
\eqa
respectively.
Since we consider a weakly-interacting Bose gas, the right-hand sides of
Eqs.~(\ref{corr1})-(\ref{corr2}) are small and we can expand the
exponents in Eq.~(\ref{1}) to first order. Combining this with
Eqs.~(\ref{dep1})-(\ref{dep2}), we finally obtain
\bqa
{n-n_{\rm c}\over n}&=&{1\over4\pi}T^{\rm 2B}(-2\mu)\;,
\eqa
in two dimensions and
\bqa
{n-n_{\rm c}\over n}&=&{32\over3}\sqrt{na^3\over\pi}\;,
\eqa
in three dimensions.
These results are in perfect agreement with the results
obtained from Popov theory.

We would like to point out that none of the other results presented in
Ref.~\cite{paper} are affected
by the above, since the theory is always applied
in a regime where the gas parameter is small and the subtlety discussed here
can be safely ignored.

\section*{Acknowledgments} The authors would like to thank Stefano
Giorgini for a useful discussion that led us to this observation.

\end{document}